\documentclass[
reprint,
bibnotes,
amsmath,amssymb,
aps,
]{revtex4-1}

\usepackage{graphicx}
\usepackage{dcolumn}
\usepackage{bm}
\usepackage{hyperref}
\usepackage[version=4]{mhchem}    
\usepackage{csquotes}
\hypersetup{colorlinks=true,linkcolor=red,citecolor=red}
\usepackage{float}
\usepackage{xspace}

\usepackage{bbding} 

\usepackage[dvipsnames]{xcolor}

\newcommand{\sco}{SrCrO$_3$\xspace}

\begin{document}

\title{Evidence for Jahn-Teller-driven metal-insulator transition in strained SrCrO$_3$ from first principles calculations}

\author{Alberto Carta}
\author{Claude Ederer}
\affiliation{Materials Theory, ETH Z\"urich, Wolfgang-Pauli-Strasse 27, 8093 Z\"urich, Switzerland}

\date{\today}

\begin{abstract}
Using density-functional theory (DFT) and its extension to DFT+$U$, we propose a possible scenario for a strain-induced metal-insulator transition which has been reported recently in thin films of \sco. The metal-insulator transition involves the emergence of a Jahn-Teller (JT) distortion similar to the case of the related rare-earth vanadates, which also exhibit a nominal $d^2$ occupation of the transition metal cation.
Our calculations indicate that, for realistic values of the Hubbard $U$ parameter, the unstrained system exhibts a C-type antiferromagnetically ordered ground state, that is already rather close to a JT instability. However, the emergence of the JT distortion is disfavored by the large energetic overlap of the $d_{xz}$/$d_{yz}$ band with the lower lying $d_{xy}$ band. Tensile epitaxial strain lowers the energy of the $d_{xy}$ band relative to $d_{xz}$/$d_{yz}$ and thus brings the system closer to the nominal filling of $d_{xy}^1(d_{xz}d_{yz})^1$.
The JT distortion then lifts the degeneracy between the $d_{xz}$ and $d_{yz}$ orbitals and thus allows to open up a gap in the electronic band structure.
\end{abstract}

\maketitle

\section{\label{sec:Intro}Introduction}

The alkaline earth chromates, $A$CrO$_3$, where $A$ can be Ca, Sr, or Ba, are a relatively unexplored class of perovskite-structured transition metal (TM) oxides where the Cr cation is in a peculiar tetravalent oxidation state (\ce{Cr^{4+}}) with a nominal $d^2$ valence electron configuration. 
While BaCrO$_3$, which has only been stabilized in epitaxial thin films~\cite{Zhu2013_BCO}, appears to be a Mott insulator~\cite{Giovannetti2014_BCO_JT}, the systems with $A$=Ca and Sr are categorized to be at the border between localized and itinerant electron behavior~\cite{Zhou2006_NM_insulator,Streltsov2008_band_localized_magnetism, Komarek2011_experimental_c_type,Long2011_itinerant_localized_crossover}. 
Both materials are reported to be rare examples of antiferromagnetic metallic oxides, and, furthermore, SrCrO$_3$ has been suggested as possible realization of a \emph{Hund's metal}, where strong correlations and significant quasiparticle renormalization are caused by the Hund's coupling~\cite{deMedici2011_Janus, Georges2013_strong_corr_Hund}.

Even though the first successful synthesis of \sco has already been reported in 1967~\cite{Chamberland1967_first_experiment}, its basic properties are still not clearly established, with various apparently contradictory experimental observations.
This is partially due to the difficulty to stabilize the Cr$^{4+}$ charge state within an octahedral environment, which requires high pressure and high temperature during bulk synthesis~\cite{Chamberland1967_first_experiment,Zhou2006_NM_insulator,OrtegaSanMartin2007_partial_OO_experimental}. 
Nevertheless, all experiments report a cubic perovskite structure and a paramagnetic susceptibility above $\sim$70-100\,K, whereby the paramagnetism cannot be well described by either an itinerant Pauli nor by a Curie-Weiss local moment picture. 
However, while some works report metallic conductivity~\cite{Chamberland1967_first_experiment, Williams2006_metallic_AFM, Komarek2011_experimental_c_type}, others observe insulating/semiconducting behavior~\cite{Zhou2006_NM_insulator, Long2011_itinerant_localized_crossover, Cao2015_paramag_semicond}, albeit with a possible transition to metallicity under pressure~\cite{Zhou2006_NM_insulator, Long2011_itinerant_localized_crossover}. 
Furthermore, a coexisting tetragonal antiferromagnetically (C-type) ordered phase has been reported at low temperatures~\cite{OrtegaSanMartin2007_partial_OO_experimental, Komarek2011_experimental_c_type}.
More recent studies on epitaxially grown thin films also find metallicity as well as indications for an antiferromagnetic (AFM) ground state~\cite{Zhang2015_metallic_AFM}. 
Very recently, a transition from metallic to insulating behavior has been reported in thin films under tensile epitaxial strain~\cite{Bertino2021_MIT_strain}.

\begin{figure}
   \centering
   \includegraphics[width=\columnwidth]{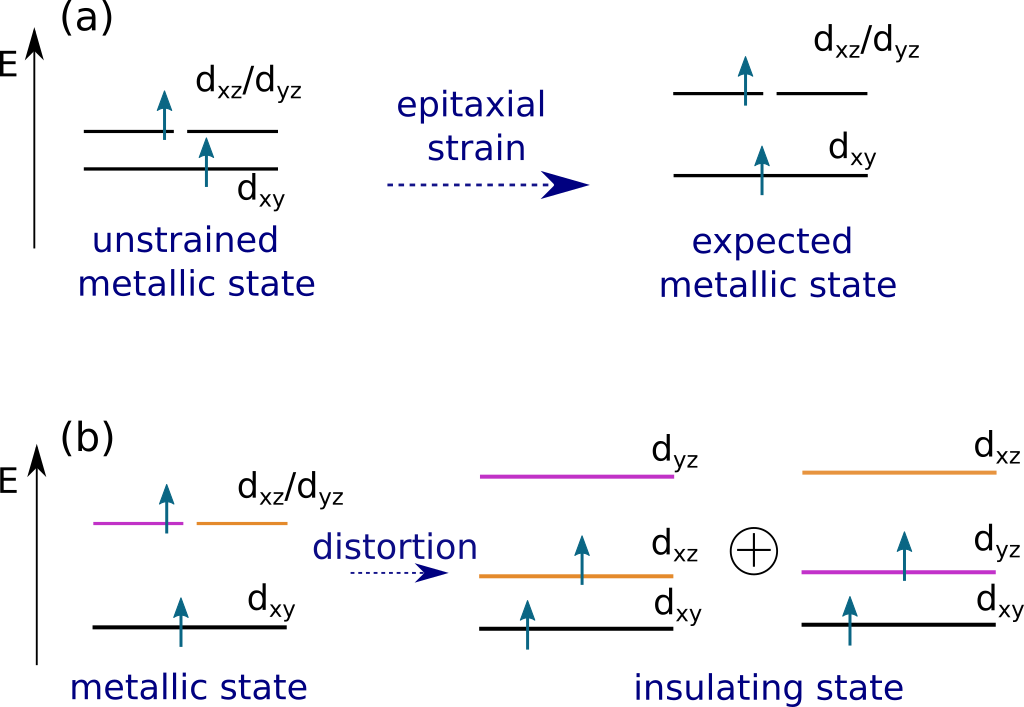}
   \caption{Effect of \textbf{(a)} epitaxial strain and \textbf{(b)} Jahn-Teller (JT) distortion on the relative energies of the $t_{2g}$ states and their occupation for a $d^2$ cation and C-type AFM order. There is a small splitting between $d_{xy}$ and $d_{xz}/d_{yz}$ states already in the unstrained metallic state, due to the C-type AFM order. \textbf{(a)} Applying tensile epitaxial strain in the $xy$ plane increases this splitting but conserves the degeneracy and partial filling of $d_{xz}$ and $d_{yz}$ levels. \textbf{(b)} In contrast, the JT distortion lifts this degeneracy, results in an alternating occupation of either $d_{xz}$ or $d_{yz}$, and the possible emergence of an insulating state.}
   \label{fig:levels}
\end{figure}

A C-type AFM metallic ground state for \sco is also supported by first principles calculations using density functional theory (DFT)~\cite{Lee2009_DFTU_OO, Qian2011_weak_correlations, Zhang2015_metallic_AFM}.
As discussed by Qian \textit{et al.}~\cite{Qian2011_weak_correlations}, the C-type magnetic order, which corresponds to a wave-vector of $(\tfrac{1}{2}, \tfrac{1}{2}, 0)$ (in units of the reciprocal lattice vectors), breaks the cubic symmetry and leads to a tetragonal distortion of the unit cell with $c/a<1$. This also results in a splitting of the Cr-$t_{2g}$ orbitals and a preferential occupation of $d_{xy}$ compared to $d_{xz}$ and $d_{yz}$ orbitals, with the latter two remaining two-fold degenerate (see Fig.~\ref{fig:levels}).
Incorporating a Hubbard $+U$ correction in the calculation, to better account for the potentially strong Coulomb interaction within the rather localized $d$ orbitals, effectively enhances the orbital splitting, eventually resulting in an orbitally polarized $d_{xy}^1(d_{xz}d_{yz})^1$ configuration for sufficiently large $U$~\cite{Lee2009_DFTU_OO}. However, the system still remains metallic due to the degeneracy of the $d_{xz}$/$d_{yz}$ orbitals.

We note that in earlier papers (e.g., ~\cite{OrtegaSanMartin2007_partial_OO_experimental}) the orbital order has been assumed to drive the structural distortion from cubic to tetragonal. However, the DFT results by Qian \textit{et al.}~\cite{Qian2011_weak_correlations} (and also our own results presented in the following), suggest that it might be more appropriate to consider the magnetic order as the main driving force, which then results in both the tetragonal distortion and the orbital polarization. Nevertheless, the coupling between orbital polarization and tetragonality could potentially also play a role in stabilizing the C-AFM state over other possible AFM wave-vectors. 

It was also reported that an insulating state with ``reversed'' orbital occupations, according to $d_{xy}^0(d_{xz}d_{yz})^2$, can be stabilized in DFT+$U$ calculations for \sco, however, only for unrealistically large values of the Hubbard $U$ parameter~\cite{Qian2011_weak_correlations}.
A similar insulating state has also been found in DFT calculations for ultrathin films and superlattices, containing only 1-2 monolayers of \sco~\cite{Gupta2013_ferroelectric_film, Zhou2015_MIT_SCO_superlattices}. 
Here, the energy of the $d_{xz}/d_{yz}$ states is lowered relative to the $d_{xy}$ orbital due to a missing apical oxygen at the film surface~\cite{Gupta2013_ferroelectric_film}, or due to a polar distortion in the superlattices~\cite{Zhou2015_MIT_SCO_superlattices}. 

In a recent preprint~\cite{Bertino2021_MIT_strain}, Bertino~\textit{et al.} demonstrate that it is possible to tune the resistivity of \sco thin films by imposing tensile epitaxial strain on the material. \sco films epitaxially grown on substrates with a small lattice mismatch remain metallic, while highly strained films grown on substrates with a larger lattice constant clearly show insulating behavior~\cite{Bertino2021_MIT_strain}.
In order to explain their findings, Bertino~\textit{et al.} resort to the aforementioned $d_{xy}^0(d_{xz}d_{yz})^2$ electron configuration resulting from a potential reordering of the $t_{2g}$ levels.
However, we note that tensile strain is expected to promote exactly the opposite trend, i.e., an energy lowering of the $d_{xy}$ orbital relative to $d_{xz}/d_{yz}$ (see, e.g., \cite{Sclauzero2016_MIT_strain_perovskites} and Fig.~\ref{fig:levels}a). Furthermore, it is unlikely that the effects promoting the reversed splitting of the $t_{2g}$ levels in ultrathin films and superlattices~\cite{Gupta2013_ferroelectric_film, Zhou2015_MIT_SCO_superlattices} will survive in \sco films of several nm thickness.

Here, we therefore suggest a different scenario for a metal-insulator transition (MIT) in \sco under tensile epitaxial strain. We show that such strain can stabilize a Jahn-Teller (JT) distortion, very similar to that observed in the closely related series of rare earth vanadates~\cite{Miyasaka2003, Reehuis2006_neutron_vanadates_0, Sage2007_vanadates_experimental, Martinez-Lope2008_RVOs, Reehuis2011_neutron_vanadates_1, Varignon2015_main_vanadates}, where the V$^{3+}$ cation has the same $d^2$ electron configuration as the Cr$^{4+}$ cation in \sco. The JT distortion lifts the degeneracy of the $d_{xz}$/$d_{yz}$ orbitals (see Fig.~\ref{fig:levels}b) and thus enables the opening of a band-gap. Our results, obtained from electronic structure calculations within DFT+$U$, show that the JT distortion can also be stabilized in the unstrained case for a sufficiently strong Coulomb repulsion between the $d$ electrons, parameterized by the Hubbard $U$. Applying  tensile epitaxial strain lowers the $U$ value required to stabilize a nonzero JT distortion, and therefore can drive a MIT in \sco.

The remainder of this paper is organized as follows. We first describe the JT modes that we investigate in our calculations (Sec.~\ref{sec:JT}), followed by a description of our computational method (Sec.~\ref{sec:Computational_details}). We then start by establishing the ground state properties of unstrained \sco as function of the Hubbard $U$ parameter (Sec.~\ref{sec:SCO}), before assessing the effect of tensile epitaxial strain on the stability of the JT distortion and the electronic structure of \sco (Sec.~\ref{sec:JT_SCO}). Finally, in Sec.~\ref{sec:summary}, we summarize our main conclusions.

\section{\label{sec:Methods} Method}

\subsection{\label{sec:JT} Jahn-Teller distorted structure}

As already described in the introduction, Sec.~\ref{sec:Intro}, an insulating state in SrCrO$_3$ can potentially be achieved by removing the degeneracy between the $d_{xz}$ and $d_{yz}$ orbital. This leads to half-filling of the corresponding orbital, which facilitates the opening of an energy gap either through magnetic ordering, leading to an additional local spin splitting, or via a Mott transition (or both).
This can be observed, e.g., in the closely related series of rare earth vanadates, $R$VO$_3$~\cite{Miyasaka2003, Reehuis2006_neutron_vanadates_0, Sage2007_vanadates_experimental, Martinez-Lope2008_RVOs, Reehuis2011_neutron_vanadates_1, Varignon2015_main_vanadates}, where the $V^{3+}$ cation exhibits the same $d^2$ electron configuration as the Cr$^{4+}$ cation in \sco.

\begin{figure}
   \centering
   \includegraphics[width=0.7\columnwidth]{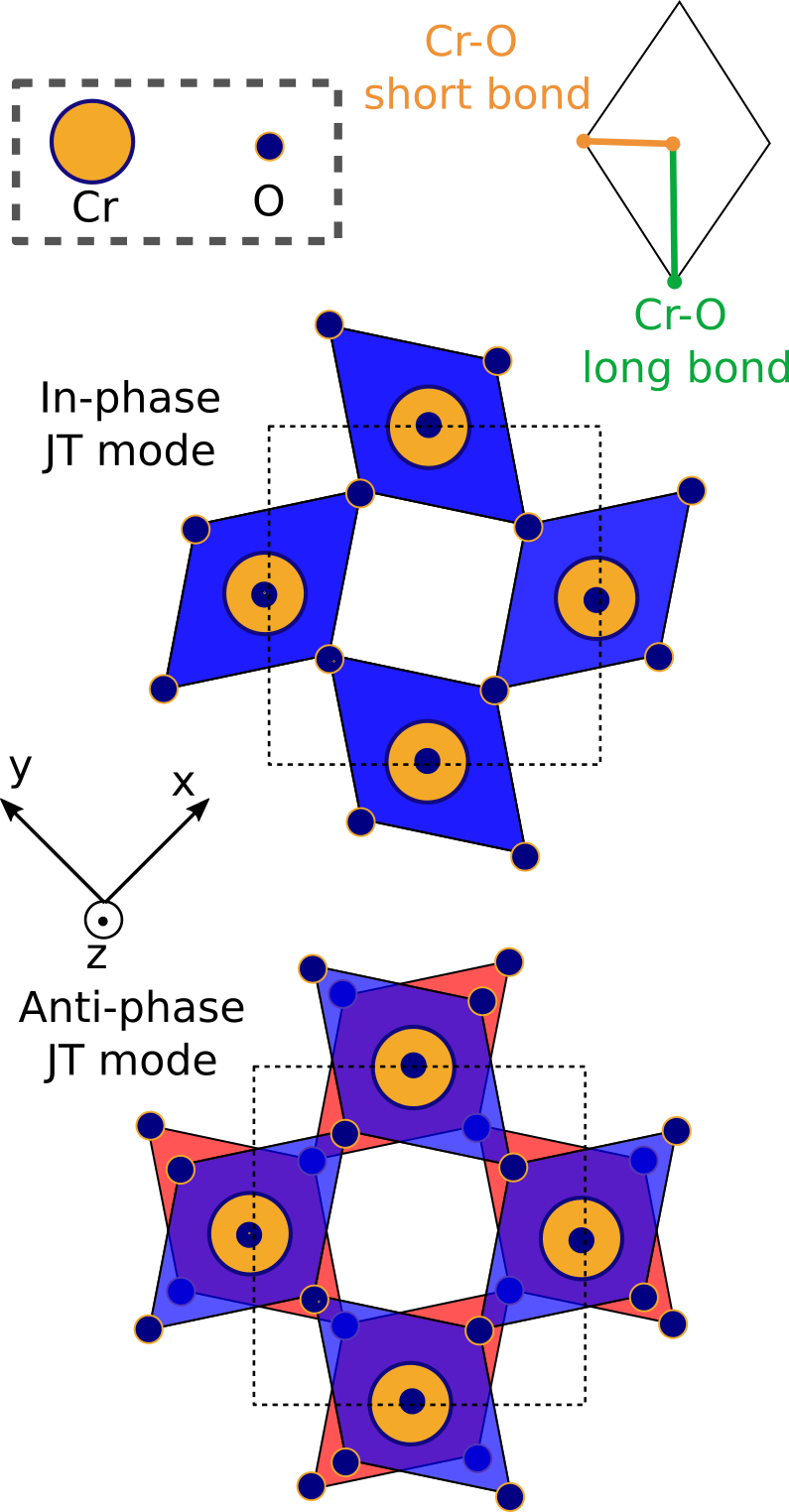}
   \caption{Orthographic projections of the in-phase $M_3^+$ (upper part) and anti-phase $R_3^-$ (lower part) JT modes considered in this work. Oxygen octahedra around the Cr atoms are marked by blue and red shading. Both modes exhibit a staggered arrangement with alternating long and short octahedral axes within the $x$-$y$ plane, but differ in the stacking along $z$.  
   The Cr-O bond distances used to quantify the amplitude of both modes (Eq.~\eqref{eq:JT}) are depicted in the top right, and the $( \sqrt{2} \times \sqrt{2} \times 2)$ supercell is indicated by the black dotted line.}
   \label{fig:JT_modes}
\end{figure}

At room temperature, the rare-earth vanadates exhibit an orthorhombically distorted perosvkite structure with space group $Pbmn$~\cite{Martinez-Lope2008_RVOs}. At low temperatures, various structural transitions can be observed, which involve two different JT modes (see Fig. \ref{fig:JT_modes}): a JT ``in-phase'' mode (irrep: $M_3^+$), which conserves the  $Pbnm$ symmetry, and a corresponding ``anti-phase'' mode (irrep: $R_3^-$), which further lowers the symmetry to $P2_1/a$~\cite{Varignon2015_main_vanadates,Sage2007_vanadates_experimental}. Both modes exhibit the same staggered arrangement of JT-distorted oxygen octahedra within an individual $(001)$-type plane, with alternating long and short O-TM-O distances along the $x$ and $y$ directions. In the in-phase $M_3^+$ mode, identical planes are stacked on top of each other along the $z$ direction, whereas in the anti-phase $R_3^-$ mode these planes are stacked with alternating orientation of long and short axes (see Fig.~\ref{fig:JT_modes}). 
Both structural distortions create a local imbalance in the occupation of the $d_{xz}$ and $d_{yz}$ orbitals on the TM sites, leading to orbital order (OO) with different ordering vector. 
The OO is strongly coupled to a potentially coexisting AFM order in such a way that G-type AFM order, corresponding to wave vector $(\tfrac{1}{2}, \tfrac{1}{2}, \tfrac{1}{2})$, always appears in combination with C-type OO, i.e, with the in-phase JT mode corresponding to wave vector $(\tfrac{1}{2}, \tfrac{1}{2}, 0)$, while C-type AFM always appears in combination with G-type OO, i.e., with the anti-phase JT mode. 

Due to the similarity between the Cr$^{4+}$ and the V$^{3+}$ cation, it appears natural to ask whether similar physics and the corresponding structural distortions can also occur in \sco, and whether this could explain the insulating behavior observed in the thin film experiments~\cite{Bertino2021_MIT_strain}. Since insulating behavior has only been observed on substrates with a large lattice mismatch, it also raises the question about the effect of epitaxial strain on such a potential MIT in \sco.

To explore these questions, we perform DFT+$U$ calculations for strained and unstrained \sco. Thereby, we mimic the elastic boundary condition imposed on epitaxially strained thin films using periodic bulk unit cells, where we fix the lattice parameters in the $x$ and $y$ directions to values slightly larger than the corresponding relaxed bulk lattice constant, while relaxing the lattice parameter along the out-of-plane $z$ direction. Furthermore, to test for the stability of a JT distorted structure, we initialize the atomic positions according to a certain amplitude of the JT distortion, either in-phase or anti-phase, and then let the atomic positions relax, allowing them to relax back to zero JT amplitude if the distortion is energetically unfavorable.

To analyze the corresponding trends, we vary both the value of the Hubbard $U$ and we change the in-plane lattice parameter. The epitaxial strain is quantified as \mbox{$\varepsilon_{xx}=\varepsilon_{yy}=(a_\text{strained}-a)/a$}, i.e., relative to the relaxed in-plane lattice parameter $a$ (always taken for the same magnetic order and $U$ value as in the corresponding strained calculations).
The amplitude of the JT modes is quantified  as:
\begin{equation}
\label{eq:JT}
 Q_\text{JT} = \frac{d_\text{Cr-O}^\text{long}}{d_\text{Cr-O}^\text{short}} - 1
\end{equation}
where $d^\text{long}_\text{Cr-O}$ ($d^\text{short}_\text{Cr-O}$) represent the long (short) Cr-O bond-distances (see Fig.~\ref{fig:JT_modes}).

\subsection{\label{sec:Computational_details}Computational Details}

We perform all DFT+$U$ calculations using the Quantum Espresso package~\cite{Giannozzi2009_QE_general} (version 6.4.1.) in combination with ultrasoft pseudopotentials from the GBRV library~\cite{Garrity2014_GBRV_pseudos}. We treat the exchange and correlation energy within the generalized gradient approximation using the functional by Perdew, Burke, and Ernzerhof modified for solids (PBEsol)~\cite{Perdew2009_PBEsol}.
To better account for the strong electron-electron interaction within the localized $d$ orbitals on the Cr, we employ the ``$+U$'' correction as implemented in Quantum Espresso \cite{Anisimov1991_original_DFTU, Cococcioni2005_QE_DFTU}, considering only an effective interaction parameter, $U_\text{eff} = U - J$, which is equivalent to setting the Hund's coupling $J=0$. 

All calculations are performed using a ($\sqrt{2} \times \sqrt{2} \times 2$) supercell relative to the primitive cell of the ideal cubic perovskite structure, which allows to accommodate both the in-phase and anti-phase JT modes as well as different AFM orders. Specifically, we consider C-type, G-type, and A-type AFM order, corresponding to wave-vectors $(\tfrac{1}{2}, \tfrac{1}{2}, 0)$, $(\tfrac{1}{2}, \tfrac{1}{2}, \tfrac{1}{2})$, and $(0, 0, \tfrac{1}{2})$, respectively (in units of the primitive reciprocal lattice vectors of the simple perovskite structure).
The plane wave kinetic energy cutoff is set to 56\,Ry for the wavefunctions and to 624\,Ry for the charge density. Brillouin zone integrations are performed using the Marzari-Vanderbilt ``cold-smearing'' scheme \cite{Marzari1999_smearing} and a ($10 \times 10 \times 7$) \textit{k}-point mesh .

\section{\label{sec:results}Results and Discussion}

\subsection{\label{sec:SCO} Bulk properties of \sco as function of Hubbard $U$}

We start by establishing the basic structural and magnetic properties of \emph{unstrained} \sco within DFT+$U$. 
For this purpose we perform structural relaxations and total energy calculations for three different magnetic states, including A-type, C-type, and G-type AFM order (see Sec.~\ref{sec:Computational_details}).

\subsubsection{The case without JT distortion}

\begin{figure}
   \centering
   \includegraphics[width=\columnwidth]{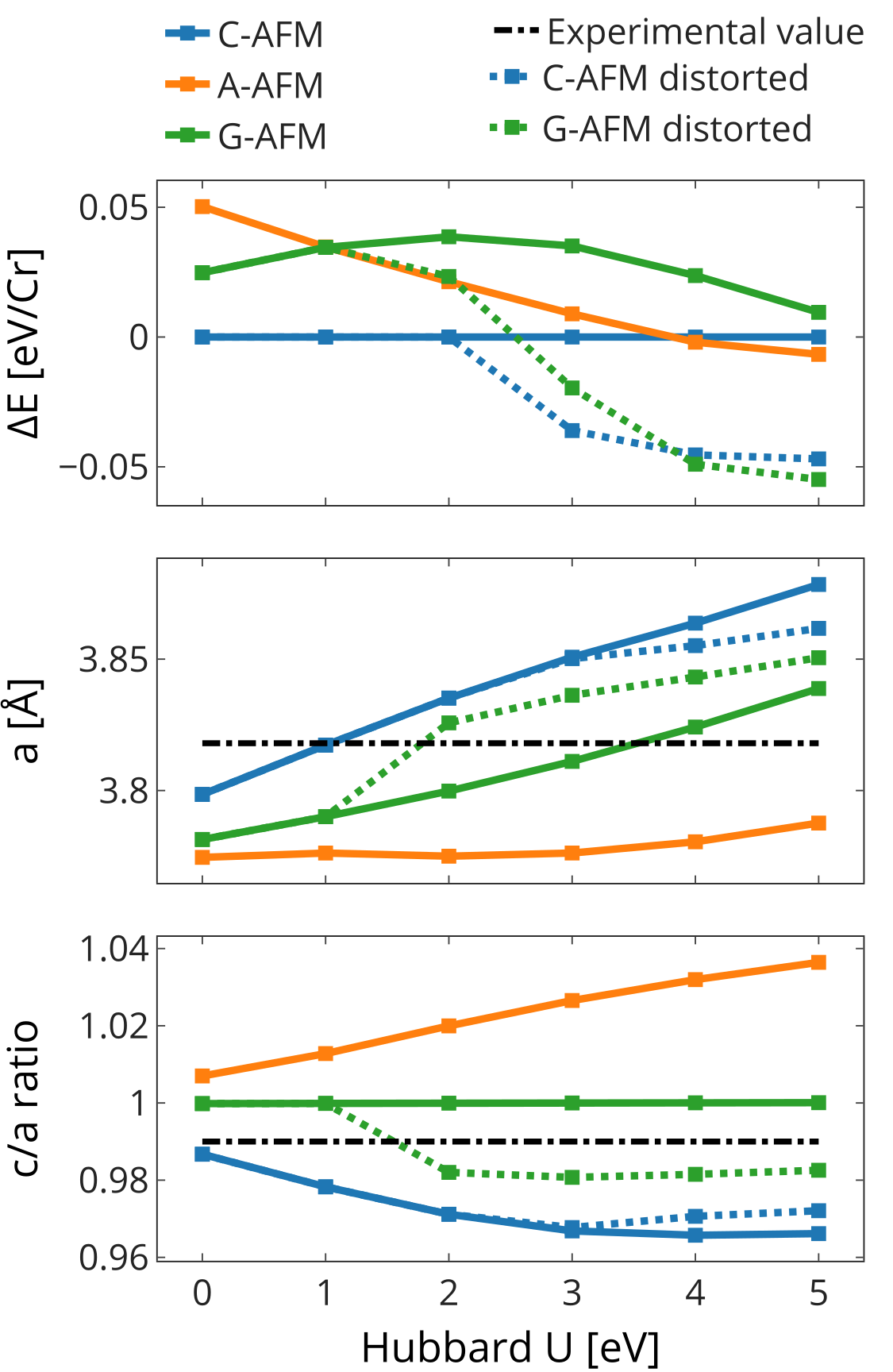}
   \caption{(a) Total energies (per Cr atom) for the different magnetic structures relative to the C-type AFM order as function of the Hubbard $U$ parameter. (b) Corresponding fully relaxed in-plane lattice constant $a$. (c) Corresponding fully relaxed $c/a$ ratio. Solid lines represent calculations where no JT distortion is allowed. Dashed lines represent calculations where the system is allowed to develop a JT distortion. The black dot-dashed lines in (b) and (c) indicate the experimental values ($a = 3.82$\,\AA{}, $c/a = 0.99$) from Ref.~\onlinecite{Komarek2011_experimental_c_type}.}
   \label{fig:en_diff_structures} 
\end{figure}

First, we discuss the case where we only allow for a tetragonal deformation of the unit cell but not for a JT distortion, i.e., we relax the volume and $c/a$ ratio of the unit cell for different $U$ values ranging from 0\,eV to 5\,eV, \emph{without} initializing a small initial JT distortion. 
The corresponding results are shown as solid lines in Fig.~\ref{fig:en_diff_structures}, where we plot the obtained energy differences between the three different magnetic states (relative to C-type AFM) as well as the resulting lattice parameters as function of the Hubbard $U$.

It can be seen that, consistently with previous DFT calculations~\cite{Lee2009_DFTU_OO,Qian2011_weak_correlations}, the C-type AFM state has the lowest energy for $U<4$\,eV, while for higher $U$ values A-type AFM eventually becomes the ground state.
For $0 < U \lesssim 2$\,eV, the corresponding relaxed in-plane lattice constant $a$ is also in rather good agreement with the experimental value $a = 3.82$\,\AA{} from Ref.~\onlinecite{Komarek2011_experimental_c_type}. Generally, $a$ increases with increasing $U$.  
Furthermore, while G-AFM order maintains the original cubic symmetry (space group $Pm\bar{3}m$), with $c/a=1$, both C- and A-type AFM reduce the symmetry to tetragonal (space group $P4/mmm$). In the case of C-AFM, this results in a contraction of the out-of-plane lattice parameter $c$, while A-AFM leads to an elongation with $c/a>1$. This supports the idea that the main driving force behind the tetragonal distortion of the material is the underlying magnetic order, as proposed by Qian \textit{et al.}~\cite{Qian2011_weak_correlations}.  
The tetragonal distortion becomes more pronounced for higher $U$ values, and overall it appears that our calculations predict a somewhat larger compression of the $c$ axis, i.e., a smaller $c/a$ ratio, for the C-AFM case compared to the value reported experimentally ($c/a = 0.99$)~\cite{Komarek2011_experimental_c_type}.

Note that we have also done analogous calculations using the local spin density approximation (LSDA) as well as the original PBE functional~\cite{Perdew1996_PBE}.  The results are qualitatively similar, except that, as expected, LSDA (PBE) results in smaller (larger) lattice constant and a $c/a$ ratio that generally deviates slightly less (more) from 1 compared to the PBEsol results. Furthermore, the transition from C- to A-AFM occurs for a slightly larger (smaller) $U$ value. Overall, the PBEsol functional appears to give lattice parameters closest to the experimental values, and we therefore only present the corresponding results here.

\subsubsection{The effects of the Jahn-Teller mode}

Next, we allow the JT distortion to develop in the material, i.e., we break the symmetry by initializing the oxygen positions with a small JT amplitude and then let the system relax to its lowest energy state. We can see in Fig.~\ref{fig:en_diff_structures} that a JT distortion develops both in the C- and G-type AFM cases for $U\geq3$\,eV and $U\geq2$\,eV, respectively. In accordance with the vanadates, and consistent with the Kugel-Khomskii model of superexchange for coupled spin and orbital degrees of freedom~\cite{Kugel/Khomskii1982}, for C-AFM only the anti-phase $R_3^-$ mode is stabilized, while for G-AFM the in-phase $M_3^+$ mode develops. We have verified that initializing the anti-phase mode in combination with G-AFM or the in-phase mode with C-AFM always results in the system relaxing back to the corresponding undistorted structure. No JT distortion can be stabilized for A-AFM, consistent with the fact that in this case the magnetic order and the resulting elongation of the $c$ axis lowers the $d_{xz}$ and $d_{yz}$ levels relative to $d_{xy}$, which results in a non-JT-active configuration with two electrons in the lower-lying two-fold degenerate orbitals.

As already noted, a slightly lower $U$ value is necessary to stabilize the JT distortion for G-AFM compared to the C-AFM case. Furthermore, the JT-distorted G-AFM state becomes the ground state of the system for $U\geq4$\,eV.
The JT distortion lowers the symmetry in the G-AFM case from cubic to tetragonal (space group $P4/mbm$), which results in an elongation of the in-plane relative to the out-of-plane lattice parameter and thus $c/a<1$. For C-AFM, the JT distortion does not have a strong effect on the lattice parameters. 

We note that previous calculations of screened interaction parameters for \sco using the constrained random phase approximation have obtained values of $U=2.9$\,eV and $J=0.85$\,eV~\cite{Vaugier2012_estimation_U}, resulting in an effective interaction parameter $U_\text{eff} \approx 2$\,eV.
Thus, we consider a $U$ of around 2\,eV as the most realistic regime for \sco. This, according to our calculations, results in an undistorted C-AFM ground state, which, however, is rather close to a JT instability.

\subsection{\label{sec:JT_SCO} \sco under tensile epitaxial strain}

As described in the introduction, Sec.~\ref{sec:Intro}, a transition to insulating behavior has recently been observed in thin films grown on substrates with a larger lattice constant than bulk \sco~\cite{Bertino2021_MIT_strain}. Therefore, we now investigate how tensile epitaxial strain affects the stability of the JT distortion, and also analyze the corresponding changes in the electronic structure of \sco. To this end, we perform calculations where we fix the in-plane lattice parameter $a$ but relax all atomic positions as well as the out-of-plane lattice parameter $c$, as described in Sec.~\ref{sec:JT}. All calculations presented in this section are performed for the C-type AFM order, which is the ground state of the system within the range of $U$ values that we are considering.

\subsubsection{Basic phase diagram under strain}

\begin{figure}
   \centering
   \includegraphics[width=\columnwidth]{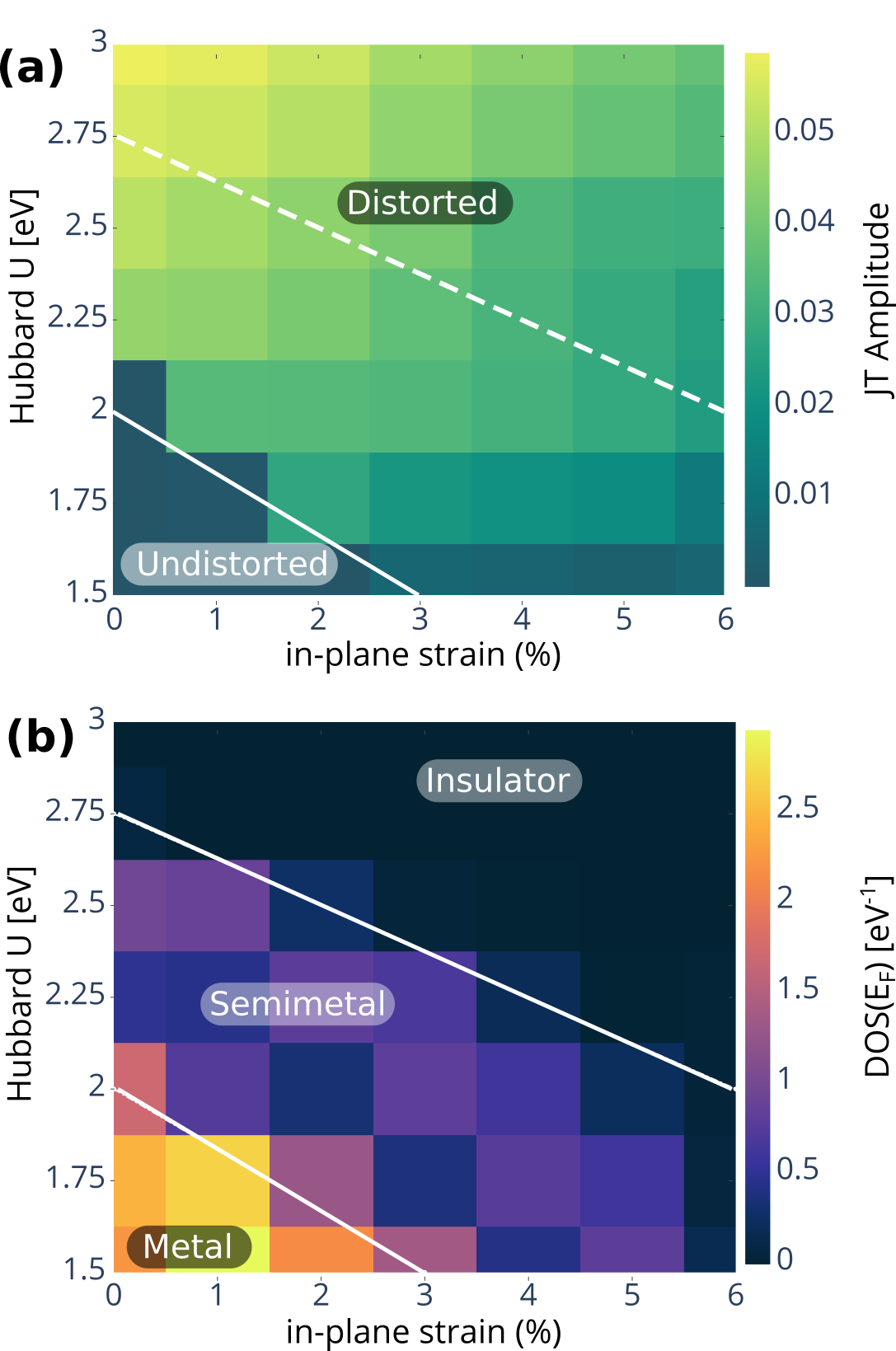}
   \caption{Evolution of \textbf{(a)} the Jahn-Teller (JT) amplitude and \textbf{(b)} the density of states at the Fermi level, DOS($E_\text{F}$), as a function of epitaxial strain and the Hubbard $U$ parameter. The white lines are a visual aid to distinguish distorted and undistorted regions in (a) and metallic, semimetallic, and insulating regions in (b). See the main text for more details. } 
   \label{fig:phase_diagram}
\end{figure}

The relaxed amplitude of the JT distortion as function of the Hubbard $U$ and 
in-plane strain (measured relative to the relaxed bulk lattice constant for each $U$) is shown in Fig.~\ref{fig:phase_diagram}(a). Note that in order to identify clear trends, we consider strain values of up to 6\,\%, even though such high strains are unlikely to be achieved in realistic thin film samples. 
For zero in-plane strain, the JT mode emerges for $U \geq 2.25$\,eV, consistent with the results presented in Fig.~\ref{fig:en_diff_structures}, and increases in amplitude for increasing $U$. For increasing strain, the minimal $U$ value required to obtain a nonzero JT distortion decreases more or less linearly. Thus, tensile strain appears to favor the JT distortion. On the other hand, one can also see that for a given $U$, once the JT distortion appears, the amplitude of the JT mode actually decreases with increasing strain. Thus, there seem to be two opposing tendencies, which we will discuss in more detail further below.

Fig.~\ref{fig:phase_diagram}(b) shows the value of the total density of states (DOS) at the Fermi level, $E_\text{F}$, as function of $U$ and in-plane strain. It can be seen that in the region where the JT distortion is zero, the system is clearly metallic, with a high density of states at $E_\text{F}$. For large $U$, the density of states at $E_\text{F}$ is zero, indicating the opening of an insulating gap. In between, there is an intermediate region where the system already exhibits a JT distortion, but is still metallic, albeit with a significantly reduced density of states at the Fermi level. This region is labeled as ``semimetal'' in Fig.~\ref{fig:phase_diagram}(b), for reasons explained in the following. 

\begin{figure*}
\centering
	\includegraphics[width=\textwidth]{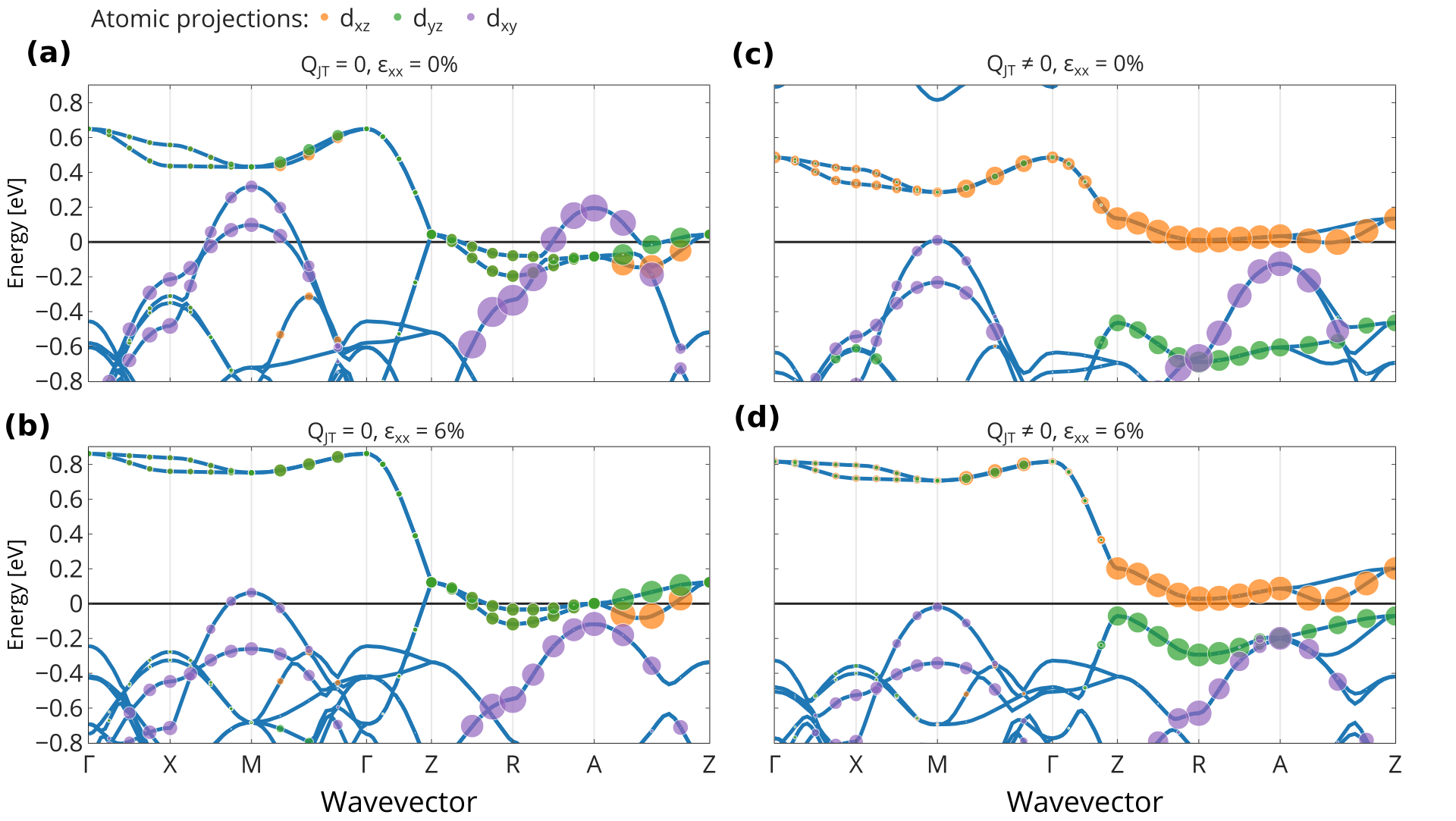} 
	\caption{Bandstructure of \sco obtained for $U=2.5$\,eV with [(c) and (d)] and without [(a) and (b)] JT distortion $Q_\text{JT}$, both for the  unstrained case [(a) and (c)] and under a tensile epitaxial strain of $\varepsilon_{xx}=6$\,\% [(b) and (d)]. The orbital character, projected on the $t_{2g}$ orbitals on one of the Cr atoms, is indicated by the colored markers, with the size of the marker representing the weight of the corresponding orbital in the Bloch function.
	The Fermi level is indicated by the horizontal black line, and $k$-point labels correspond to the primitive tetragonal Brillouin zone of the $\sqrt{2} \times \sqrt{2} \times 2$ cell.}
	\label{fig:bands} 
\end{figure*}

\subsubsection{Analysis of band structure}

To understand the existence of the JT distortion in the ``semimetallic'' phase, and the apparently opposing tendencies under strain (emergence of JT distortion but then decreasing amplitude for increasing strain), we now compare the band structure in the energy region around the Fermi level for the cases with and without JT distortion, both for zero strain and for an in-plane strain of 6\,\%. 
Fig.~\ref{fig:bands} shows the corresponding band-structures for $U=2.5$\,eV. 
The orbital character of the individual bands is also indicated, by projecting the Bloch states on $t_{2g}$ orbitals centered on one of the Cr atoms. 

We first discuss the unstrained case without JT distortion, shown in Fig.~\ref{fig:bands}(a). In this case, one recognizes a $d_{xy}$-dominated band located mostly around and below $E_\text{F}$ (note that due to the presence of several Cr atoms within our unit cell there is more than one ``$d_{xy}$-state'' per $k$-point), and a slighly higher-lying $d_{xz}$/$d_{yz}$-dominated band. This essentially corresponds to the schematic level diagram shown in Fig.~\ref{fig:levels}(a). 
However, due to the large bandwidth, the two sets of bands overlap, and the $d_{xy}$-band is not completely filled, but crosses the Fermi level, rendering the material metallic.

As can be seen in Fig.~\ref{fig:bands}(c), the JT distortion splits the $d_{xz}$/$d_{yz}$-band at the Z-point and along \mbox{Z-R-A-Z}, thereby completely disconnecting the higher-lying band (with dominant $d_{yz}$ character on the selected Cr atom) from all lower-lying bands. Note that for the selected Cr atom, the short (long) Cr-O bond in the JT-distorted structure is oriented along $y$ ($x$), and thus the splitting of the $d_{xz}$/$d_{yz}$ band is consistent with the splitting expected from Fig.~\ref{fig:levels}(b).  
However, for the chosen value of $U=2.5$\,eV, the bottom of the higher-lying $d_{xz}$ band (along the line A-Z) still has an indirect overlap with the top of the $d_{xy}$ band (at M-point) and therefore both bands contribute to the small DOS at the Fermi level (see Fig.~\ref{fig:en_diff_structures}(b)). We thus refer to this regime as ``semimetallic''.
Furthermore, we note that, even though there is no global energy gap in the DOS, the splitting of the $d_{xz}$/$d_{yz}$ band and the resulting ``disconnected'' bandstructure apparently leads to an energy gain that is sufficient to stabilize the JT distortion for $U=2.5$\,eV (see Fig.~\ref{fig:phase_diagram}).

From the comparison of Fig~.\ref{fig:bands}(a) and (c), one can see that tensile strain pushes the $d_{xy}$-band to lower energies relative to the $d_{xz}$/$d_{yz}$-band, again consistent with the expected effect on the corresponding energy levels shown in Fig.~\ref{fig:levels}(a). This brings the system closer to the nominal occupations with a fully occupied $d_{xy}$-band and one electron (per Cr) in the $d_{xz}$/$d_{yz}$-band, and thus facilitates the stabilization of a JT distortion for a smaller $U$ value. 

On the other hand, tensile strain also leads to a contraction of the lattice parameter along $z$ due to the Poisson effect, which in turn promotes electron hopping along this direction and thus increases the bandwidth of the $d_{xz}$/$d_{yz}$-band. This increase of the bandwidth with increasing tensile strain reduces the energy gain from the JT splitting and is therefore responsible for the reduction of the JT amplitude with strain (for fixed $U$ value). The smaller JT amplitude is also reflected by the smaller splitting  between the $d_{xz}$ and $d_{yz}$ bands along Z-R-A-Z in Fig.~\ref{fig:bands}(d) compared to that in Fig.~\ref{fig:bands}(c). Nevertheless, due to the smaller overlap with the $d_{xy}$ band in the strained case, a smaller $d_{xz}$/$d_{yz}$ splitting is sufficient to fully open a gap in the band-structure, and thus the JT-distorted system becomes insulating.

\subsubsection{\label{sec:H_eff} Evolution of the orbital polarization}

\begin{figure}
   \centering
   \includegraphics[width=\columnwidth]{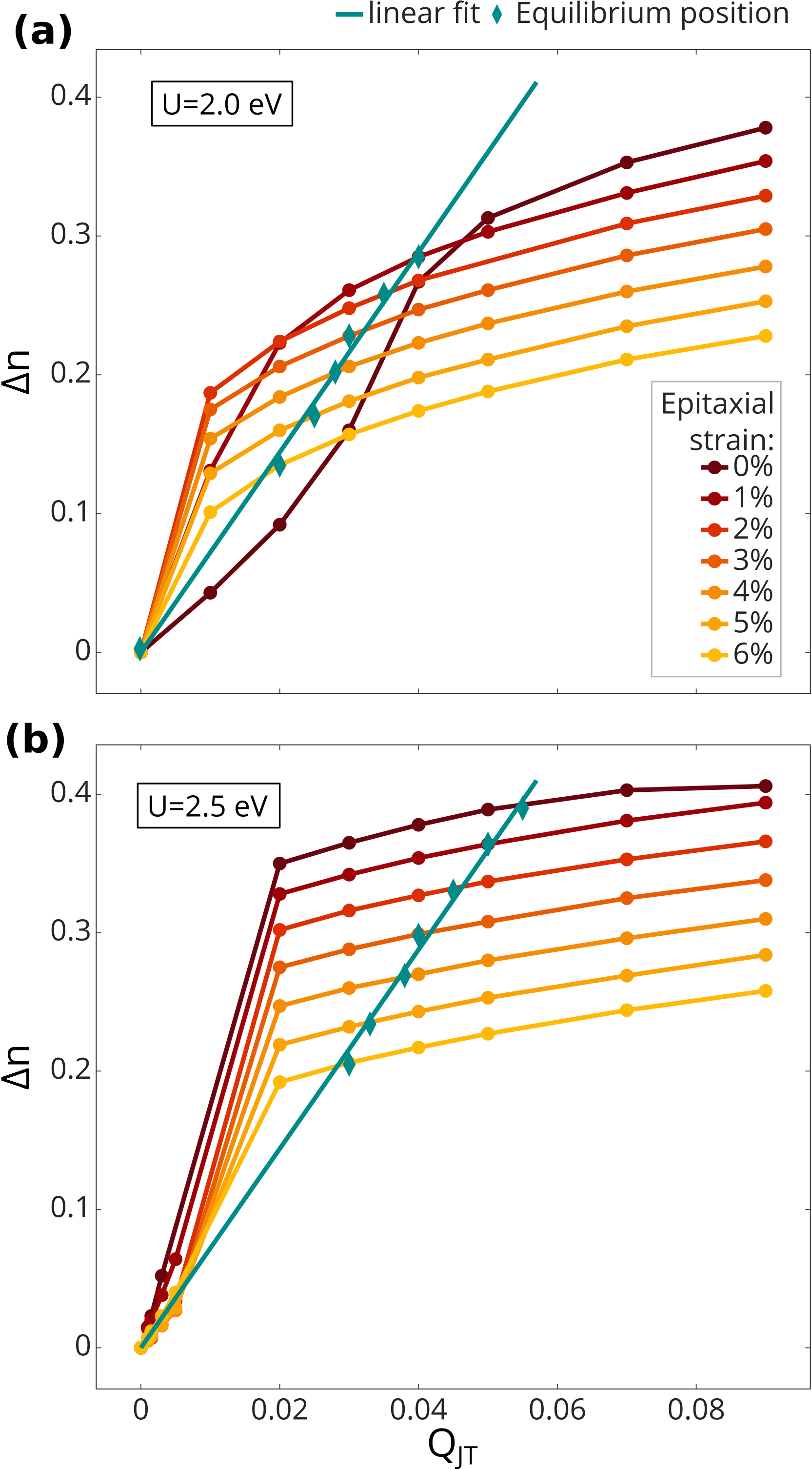}
   \caption{Evolution of the local orbital polarization $\Delta n$ as function of the amplitude $Q_\text{JT}$ of the JT distortion at different strain levels, obtained for (a) $U=2.0$\,eV and (b) $U=2.5$\,eV. The values for $\Delta n$ and $Q_\text{JT}$ that correspond to relaxed atomic positions for each strain are marked by the green diamonds, and the green line is a linear fit to these data points.} 
   \label{fig:landau_all}
\end{figure}

In order to further analyze the interplay between structural and electronic degrees of freedom, we perform calculations where we manually vary the amplitude of the JT distortion, $Q_{\text{JT}}$, for fixed strain, and then monitor the evolution of the local orbital polarization, $\Delta n$, defined as the absolute value of the occupation difference between the $d_{xz}$ and $d_{yz}$ orbitals on each Cr.
The results are shown in Fig.~\ref{fig:landau_all} for two different $U$ values.

We first discuss the case with $U=2.5$\,eV, Fig.~\ref{fig:landau_all}(b), for which the JT mode is energetically preferred at all strain levels. 
For small JT distortion ($Q_\text{JT} < 0.01$), one observes a linear regime, where $\Delta n$ is proportional to $Q_\text{JT}$, and the system exhibits a metallic density of states. 
On increasing $Q_\text{JT}$, the response seems to abruptly jump to a larger value of the orbital polarization, which then increases only slightly on further increase of $Q_\text{JT}$. This behavior corresponds to the semimetallic/insulating regime. 
The amount of orbital polarization that is achieved in this regime is decreasing with increasing strain, consistent with the smaller splitting and increased bandwidth of the $d_{xz}/d_{yz}$ band seen in Fig.~\ref{fig:bands}(d) compared to  Fig.~\ref{fig:bands}(c).
Similar behavior is also observed for $U=2.0$\,eV, except for the corresponding unstrained case, where the JT mode is not energetically stable. Here, the initial linear response is weaker and the transition to the ``saturation regime'' is much more gradual and smoothed out.

We note that in the transition region between the two regimes, the calculations are rather difficult to converge. This is likely due to the presence of both metallic and semimetallic/insulating solutions. Thus, we do not attempt to fully resolve this transition region.
We also note that the maximum orbital polarization of around $\Delta n = 0.4$ is smaller than the nominal value of 1 expected from the simple level diagram shown in Fig.~\ref{fig:levels}. This is due to hybridization between the Cr $d$ orbitals and the $p$ orbitals of the surrounding oxygen ligands.

Most importantly, one can see that the points corresponding to the equilibrium JT amplitude for each case, which are marked by green diamonds in Fig.~\ref{fig:landau_all}, can be nicely fitted by a straight line, which has the same slope for both $U=2.0$\,eV and $U=2.5$\,eV.
This is consistent with a simple model that only involves a linear coupling between the structural distortion and the corresponding electronic order parameter, in this case the orbital polarization~\cite{Peil2019_landau_theory_eff_model,Georgescu2021_MIT_E_landscape}:
\begin{equation}
E = E_{el}[\Delta n] -  g \frac{Q_{\text{JT}}\Delta n}{2} + K \frac{Q_{\text{JT}}^2}{2}.
    \label{eqn: total_energy}
\end{equation}
Here, the total energy $E$ is divided into three contributions: a purely electronic part $E_{el}$, which depends on the orbital polarization $\Delta n$, a linear coupling between $\Delta n$ and the associated structural distortion (in this case the JT mode amplitude $Q_{\text{JT}}$), and a term describing the stiffness of the structural distortion.
Minimizing the total energy with respect to $Q_\text{JT}$ and using the Hellman-Feynman theorem yields an equation of state for the equilibrium values of $Q_\text{JT}$ and $\Delta n$~\cite{Peil2019_landau_theory_eff_model, Georgescu2021_MIT_E_landscape}:
\begin{equation}
\frac{2K}{g} Q_{\text{JT}} = \Delta n (Q_{\text{JT}}) \quad .
    \label{eqn:saddle_point}
\end{equation}
Thus, the equilibrium values can be obtained from the crossing point of $\Delta n(Q_\text{JT})$ with the straight line \mbox{$2K/g \cdot Q_\text{JT}$}, where the slope is determined by the mode stiffness $K$ and the electron-lattice coupling constant $g$.

Even though we have not attempted to extract values for $K$ and $g$ from our DFT+$U$ calculations, it is apparent that the data shown in Fig.~\ref{fig:landau_all} is well described by this simple model, with a value for $K/g$ that is independent of $U$. Also the case without a stable JT distortion ($U=2.0$\,eV and $\varepsilon_{xx}=0$) is correctly described by this model, since the corresponding curve for $\Delta n(Q_\text{JT})$ does not intersect with the fitted line for $2K/g \cdot Q_\text{JT}$ for any value of $Q_\text{JT}>0$.

\section{Summary and Conclusions}
\label{sec:summary}

In summary, we have used DFT+$U$ calculation to investigate the ground state properties of \sco and the possible emergence of a JT distortion, which could lead to a metal-insulator transition under strain. Our calculations confirm a C-type AFM ground state (for $U<4$\,eV) and indicate that for a realistic value of $U \approx 2$\,eV, the system is very close to a JT instability. 
Tensile strain reduces the $U$ value that is required to develop a JT distortion to below 2\,eV, and thus can trigger a transition to the JT distorted structure. The emergence of the JT distortion is accompanied by a significant reduction in the density of states at the Fermi level and ultimately by the opening of a band gap.
This provides a realistic scenario to explain the strong increase in resistivity that has been observed experimentally in \sco thin film grown on substrates with a larger lattice constant than \sco~\cite{Bertino2021_MIT_strain}.

Our calculations also show that, while strain plays an important role in triggering the onset of the JT mode, further straining actually decreases the corresponding mode amplitude. This can be understood from the two opposing tendencies resulting from the lowering of the $d_{xy}$ band relative to $d_{xz}/d_{yz}$, and from the increasing bandwidth of the latter.
Furthermore, the emergence of the JT distortion and the associated orbital polarization appears to be well described by a simple model that involves a linear coupling between electronic and structural order parameters.

Interestingly, the emergence of the JT distortion does not lead to an abrupt metal-insulator transition. Instead, this transition occurs via an intermediate semimetallic phase, and for a fixed $U=2$\,eV a large strain of nearly 6\,\% is required to cross this semimetallic region. We note, however, that the strong reduction of DOS($E_\text{F}$) can potentially result in large changes in resistivity similar to a metal-insulator transition. It is also conceivable that the unstrained system is in fact already in the semimetallic JT-distorted state, corresponding, e.g., to $U=2.5$\,eV, but that the small JT distortion has so far eluded experimental detection. In this case, a much smaller strain would be sufficient to bring the system in the fully insulating region. 
It is also possible that the DFT+$U$ approach overestimates the extension of the semimetallic region or that the screening of the Hubbard interaction is changing under strain or as a result of the metal-insulator transition. If this would lead to an increase in the effective $U$ value, that would reinforce the transition to the insulating state.  
Finally, other influences, not included in our calculations, such as, e.g., varying defect concentrations or inhomogeneities, could also affect the metal-insulator transition. 

Further experimental characterization is required to resolve these remaining question, and we hope that our computational analysis will stimulate future experimental studies of \sco, both in bulk form and in thin films, or also as part of more complex heterostructures. 
Our work highlights some of the physics that is likely to be relevant in \sco and possibly also other alkaline earth chromites, and shows that these materials are indeed at the border between different structural and electronic instabilities.

\begin{acknowledgments}
This research was supported by the NCCR MARVEL, a National Centre of Competence in Research, funded by the Swiss National Science Foundation (grant number 182892), and by ETH Z\"urich.
Calculations were performed on the cluster \enquote{Euler} of ETH Z\"urich. We also thank Simon J\"ohr and Marta Gibert for insightful discussions and clarifying some of the experimental challenges.
\end{acknowledgments}

\bibliography{main}
\end{document}